\documentclass[fleqn,usenatbib,useAMS]{mnras}

\usepackage{graphicx}	
\usepackage{amsmath}	
\usepackage{amssymb}	
\usepackage{multicol}        
\usepackage{bm}		
\usepackage{pdflscape}	

\usepackage{newtxtext,newtxmath}

\usepackage[T1]{fontenc}

\DeclareRobustCommand{\VAN}[3]{#2}
\let\VANthebibliography\thebibliography
\def\thebibliography{\DeclareRobustCommand{\VAN}[3]{##3}\VANthebibliography}

\title[]{A new distance to the Brick, the dense molecular cloud $G0.253+0.016$.
   \thanks{Based on data taken at the ESO/VLT telescope, within the observing programme 0103.B-0262(A), PI: M. Zoccali}
   }
\author[M. Zoccali et al.]{
M. Zoccali,$^{1,2}$\thanks{E-mail: mzoccali@astro.puc.cl}
E. Valenti,$^{3,4}$
F. Surot,$^{5,6}$
O.A. Gonzalez,$^7$
A. Renzini$^{8}$ 
\newauthor
and A. Valenzuela Navarro$^{1,2}$ \\
\\
$^{1}$Instituto de Astrof\'{i}sica, Pontificia Universidad Cat\'{o}lica de Chile, Av. Vicu\~{n}a Mackenna 4860, Santiago , Chile \\
$^{2}$Millennium Institute of Astrophysics, Av. Vicu\~{n}a Mackenna 4860, 782-0436 Macul, Santiago, Chile \\
$^{3}$European Southern Observatory, Karl Schwarzschild\--Stra\ss e 2, D\--85748 Garching 
bei M\"{u}nchen, Germany \\
$^{4}$Excellence Cluster ORIGINS, Boltzmann\--Stra\ss e 2, D\--85748 Garching bei M\"{u}nchen, Germany \\
$^{5}$Instituto de Astrof\'{i}sica de Canarias, E\--38205, La Laguna, Tenerife, Spain\\
$^{6}$Departamento de Astrof\'{i}sica, Universidad de La Laguna, E\--38205, La Laguna Tenerife, Spain\\
$^{7}$ UK Astronomy Technology Centre, Royal Observatory, Blackford Hill, Edinburgh, EH9 3HJ, UK \\
$^{8}$ Istituto Nazionale di Astrofisica -- Osservatorio Astronomico di Padova, Vicolo dell’Osservatorio 5, I-35122 Padova, Italy \\
}

\date{Accepted XXX. Received YYY; in original form ZZZ}

\pubyear{2020}

\begin{document}
\label{firstpage}
\pagerange{\pageref{firstpage}--\pageref{lastpage}}
\maketitle

\begin{abstract}

We analyse the near infrared colour magnitude diagram of a field including the giant molecular cloud $G0.253+0.016$ (a.k.a. The Brick) observed at high spatial resolution, with HAWK-I@VLT. The distribution of red clump stars in a line of sight crossing the cloud, compared with that in a direction just beside it, and not crossing it, allow us to measure the distance of the cloud from the Sun to be 7.20, with a statistical uncertainty of $\pm$0.16 and a systematic error of $\pm$0.20 kpc . This is significantly closer than what is generally assumed, i.e., that the cloud belongs to the near side of the central molecular zone, at 60 pc from the Galactic center. This assumption was based on dynamical models of the central molecular zone, observationally constrained uniquely by the radial velocity of this and other clouds. 

Determining the true position of the Brick cloud is relevant because this is the densest cloud of the Galaxy not showing any ongoing star formation. This puts the cloud off by 1 order of magnitude from the Kennicutt-Schmidt relation between the density of the dense gas and the star formation rate. Several explanations have been proposed for this absence of star formation, most of them based on the dynamical evolution of this and other clouds, within the Galactic center region. Our result emphasizes the need to include constraints coming from stellar observations in the interpretation of our Galaxy's central molecular zone.

\end{abstract}




%

\section{Introduction}
\label{sec:intro}
 
The Central Molecular Zone (CMZ) is one of the most extreme environments in the Milky Way. 
It includes about $3-5\times10^7$ M$_\odot$ of molecular gas \citep{dahmen+98, pierce-price+00}, representing a relatively large (5$\%$) fraction of the total molecular gas of the Galaxy \citep{heyer+15, roman-duval+16, nakanishi+16}, concentrated in a small region of about 400 pc in the plane, and $<100$ pc in height perpendicular to it. 
The conditions of this gas, however, are pretty different from those of the rest of the disk. 
The temperature is on average hotter ($50-400$K), the density is larger, reaching peaks of n=$10^5 - 10^7$ cm$^{-3}$, turbulence is higher, there is a large variety of molecules and a stronger magnetic field \citep[and references therein]{battersby+20}.

Given the conditions of the gas in the CMZ, it is legitimate to compare this region with star forming disks at high redshift \citep{kruijssen+13, barnes+17}, with the great advantage that the CMZ can be resolved and studied in much greater detail. 
In this respect, according to the Kennicutt-Schmidt law \citep{kennicutt+89, kennicutt+98}, above a critical density, the cold and dense gas should form stars with an efficiency proportional to the gas surface density. 
The relation has been found to hold, with the same slope, from star forming regions in the Milky Way \citep[e.g.,]{wu+05, heiderman+10, lada+10, lada+12, evans+14} to parts of external galaxies \citep[e.g.,][]{bigiel+15, bigiel+16, usero+15, gallagher+18, querejeta+19, morselli+2020} out to unresolved galaxies \cite[e.g.]{gao+04a, gao+04b, krips+08}. 
The CMZ, however, is a notable exception. 
Although some ongoing star formation is observed in this region, its rate ($0.05-0.1$ M$_\odot$ yr$^{-1}$) is about an order of magnitude below what is expected given the amount of dense gas contained in the CMZ \citep{immer+12, lada+12, longmore+13, barnes+17, kauffmann+17}. 
Not surprisingly, there is something about the reason why some clouds form stars, and some
other do not, that we do not completely understand. 
The CMZ might provide the opportunity to clarify it, and make the link between our Galaxy and high-z ones.

The 3D structure of the CMZ is still a matter of debate. Given the impossibility to directly derive distances to gas structures,
their location has been inferred through dynamical models of the whole CMZ, assuming a state of the art Galactic potential and 
required to reproduce the observed radial velocities of the gas at different lines of sight, as traced by molecules such as CO,
CS, NH${\rm _3}$ and HNCO. The simplest and most widely adopted model proposes that most of the dense molecular clouds are arranged in a twisted ($\infty$-shaped) elliptical ring with semi-major axis of ~100 pc and semi-minor one of 60 pc \citep{molinari+11}. 
Notable giant molecular clouds (GMC) composing the ring are Sgr B2 (n=$10^7$ cm$^{-3}$), the most massive molecular cloud in the CMZ and also an active site of star formation, GMC$-$0.02$-$0.07 (the 50 km s$^-1$ cloud), GMC$-$0.13$-$0.08 (the 20 km $s^{-1}$ cloud), and GMC$0.25+0.01$ (a.k.a. G0.253+0.016, or the “Brick”).  Alternative, but not too different, interpretations propose that the ring is in fact composed by two spiral arms \citep{sofue95, sawada+04, ridley+17}, or it is an open stream instead \citep{kruijssen+13}. 
Aside from this few models, all the other investigations about the physical conditions of the gas, assume that all the clouds
are located at the distance of the Galactic center. It is important to establish where are the clouds, and what is their
orbit, in order to understand the shocks they were submitted to, and try to understand why some of them are forming stars while others do not.

In addition to the CMZ, within a radius of $\sim$300 pc from the Galactic center, there is a large concentration of stars usually called the Nuclear bulge. 
This component appears distinct from the Galactic bulge because it has a much higher surface density of stars with a clear break, and because it hosts a population of massive \-- and therefore young \-- stars \citep{dong+12}, three massive young (2-6 Myr) clusters, namely: the Nuclear cluster, Arches, and Quintuplet. 
In the same region, and roughly coincident with the CMZ, there is also the Nuclear Stellar Disk: a flat component that shows a larger and coherent rotation in the Galactic plane \citep{launhardt+02, schoenrich+15}.
Most of the mass in this component \citep[$3\times10^9$ M$_\odot$][their Fig.5]{valenti+16} is in old stars \cite{nogueras-lara+20}. 
While in the CMZ the ISM has been studied systematically for decades, its stellar component counterpart is still poorly explored because it requires the spatial resolution that only the Hubble Space Telescope or Adaptive Optics systems can provide.
Therefore, so far the study of the stellar population in the CMZ has been limited to very small regions, either centered on the massive clusters \citep{fritz+16, clarkson+12, habibi+14, hosek+15, stolte+05, liermann+12, hussmann+12, shin+16} or mapping a small area very close to the Galactic center \citep[e.g.,]{nogueras-lara+18, nogueras-lara+20}. 
A systematic high spatial resolution study of the stellar counterpart of the entire CMZ is currently still lacking. 

In an attempt to fill the gap between the sparse, small fields studied at high spatial resolution and the wide area surveys covering the whole bulge but not properly resolving the region within R$<300$ pc, we started a project aimed at mapping the Nuclear bulge with HAWK\--I at VLT.
Unfortunately, only a small fraction of the programme originally approved (0103.B-0262(A)) was completed, providing data only for the 5 highest priority innermost pointings. 
In the present paper we analyse the field containing one of the most prominent clouds in the CMZ, known as $G0.253+0.016$, or "The Brick", at coordinates ($l,b$)=($0.253^\circ,0.016^\circ$). 
In a re-analysis of far-IR and mm data, \citet{brick-Longmore+12} concluded that the Brick is a high-mass ($1.3\times10^5$ M$_\odot$), low-temperature (T$_{\rm dust}$ $\sim$20 K), high-volume, and column density 
(n$\sim 8\times 10^4 {\rm cm}^{-3}$; N$_{\rm H2}\sim 4\times 10^{23} {\rm cm}^{-2}$) molecular clump close to virial equilibrium. 
As such, it should be on the verge of forming stars, potentially a massive cluster like Arches. 
Instead, the Brick is the densest cloud in the Milky Way that is not actively forming stars. 
In the attempt of understanding the reason why this cloud departs from the Kennicutt-Schmidt law,
\citet{henshaw+19} analyzed the line-of-sight velocity distribution of the gas in this cloud, concluding that it is not really a “Brick”, i.e., not very concentrated along the line of sight but rather a complex and structured cloud whose distance spread is comparable to its physical size in the plane of the sky. 

Although the conditions of the gas in this particular cloud have been studied by different authors, its position within the Galaxy has been inferred only once, based on the assumption that it belongs to the so-called "100 pc elliptical ring", whose rotation has been modeled by \citet{molinari+11} and its axes thereby constrained. According to that model, the Brick should lie on the front side of the elliptical ring, at a distance of 60 pc (the ring semi-minor axis) from the Galactic center. If we assume that the Galactic center is at $8.2$ kpc from the Sun \citep{GravityCo+19}, then the Brick is at 8.14 kpc from us. A slighly different model has been proposed by \citet{ridley+17}, where the ring is made up by two arms of a nuclear spiral following very closely the $x_2$ orbits of the Galactic main bar. Within this model, the radial velocity of a cloud at the position of the Brick is consistent with the observed velocity of the latter, although that would also be true for a cloud located at the same distance from the Galactic center, but symmetrically behind it. The same degeneracy holds, in general, for every cloud in the CMZ.  \citet{brick-Longmore+12} attempted to put an independent constraint on the distance of the Brick, using the stars in front and around it. Using the distribution of the stars along the line\--of\--sight they found that the cloud should be at 7.4$\pm$1 kpc from us. However, because the stellar photometry used in that work was rather shallow, with large errors and few stars, they considered this measurement less reliable than that by \citet{molinari+11} (thought still compatible with it, within the errors) and throughout their analysis they assumed the cloud to be located at the Galactic center distance.

In the present paper we propose a re-determination of the Brick distance by using deep and accurate photometry of the stars in front and around it, following an approach very similar to that by \citet{brick-Longmore+12}. 
We anticipate that our results confirm that the Brick is significantly closer to us than the Galactic center, hence in contrast with the idea that it should belong to the 100-pc elliptical ring.

\section{Observation and data reduction}
\label{sec:obs}

\begin{figure}
    \centering
    \includegraphics[width=\columnwidth]{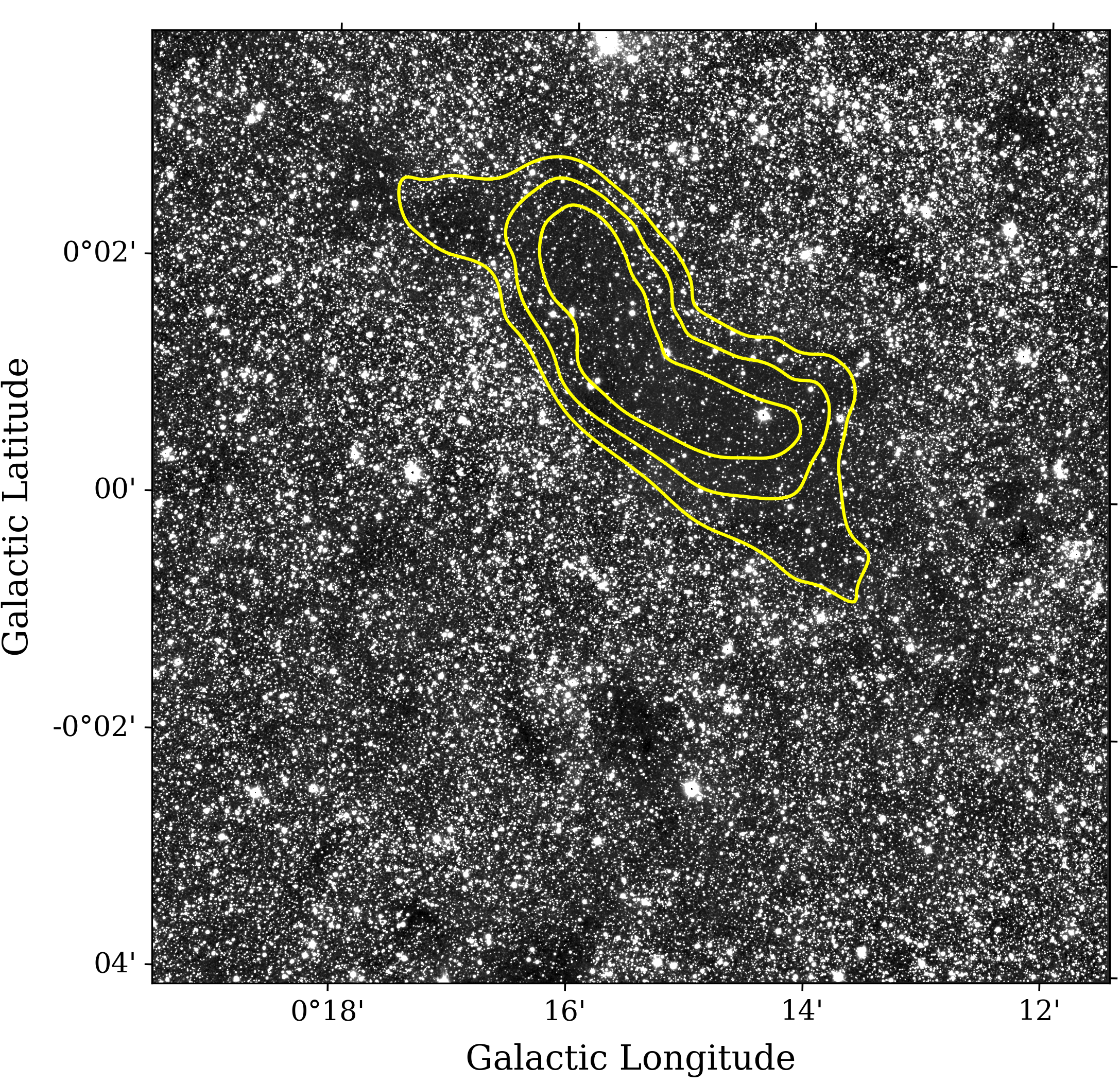}
    \caption{Final HAWK\--I $K_s$\--band tiled image where the brick clearly stands out as a region of low stellar density due to its excess extinction. Solid yellow lines refer to the Hi\--GAL column density map contours from \citet{brick-Longmore+12} (see their Fig.\,3)}
    \label{fig:hawki}
\end{figure}

In this work we use H and K$_s$ band observations of the infrared (IR) dark cloud $G0.253+0.016$ collected with HAWK\--I on the Yepun (VLT\--UT4) telescope at the ESO Paranal Observatory. HAWK\--I is a wide field ($7.5'\times 7.5'$) near\--IR imager with a pixel scale of 0.106"/px and equipped with the GRound layer Adaptive optics system Assisted by Laser (GRAAL), which provides image quality improvement (i.e., it is a seeing enhancement system).
A set of 16 and 8 exposures, each 30\,s and 20\,s long, was obtained through the H and K$_s$ pass\--band filters, respectively. 
The exposures were taken following a dithering pattern specifically designed: {\it i)} to homogeneously cover the small cross\--shape gap of $\sim 15"$ between the four detectors mosaic; and {\it ii)} for background subtraction.
The observations were taken in service mode in July 2019 (ESO period 103) under an average optical seeing of 0.7" that combined with the GRAAL correction led to a final image quality of 0.4" and 0.3" in H and K$_{\rm s}$, respectively.

The raw data have been processed by using the ESO HAWK\--I pipeline within the Reflex workflow\footnote{ftp://ftp.eso.org/pub/dfs/pipelines/instruments/hawki}, which produces final tiled (see Figure\,\ref{fig:hawki}) and stacked pawprint images for each filter. The total exposure time of the stacked images is, therefore, 480\,s and 160\,s in the H and K$_{\rm s}$ band, respectively. The Brick is clearly visible in this image as a dark cloud, almost devoid of stars, in the middle. Yellow contours are taken from the column density maps of \citet{brick-Longmore+12}, obtained within the HiGAL survey \citep{molinari+10}.

\begin{figure}
    \centering
    \includegraphics[width=\columnwidth]{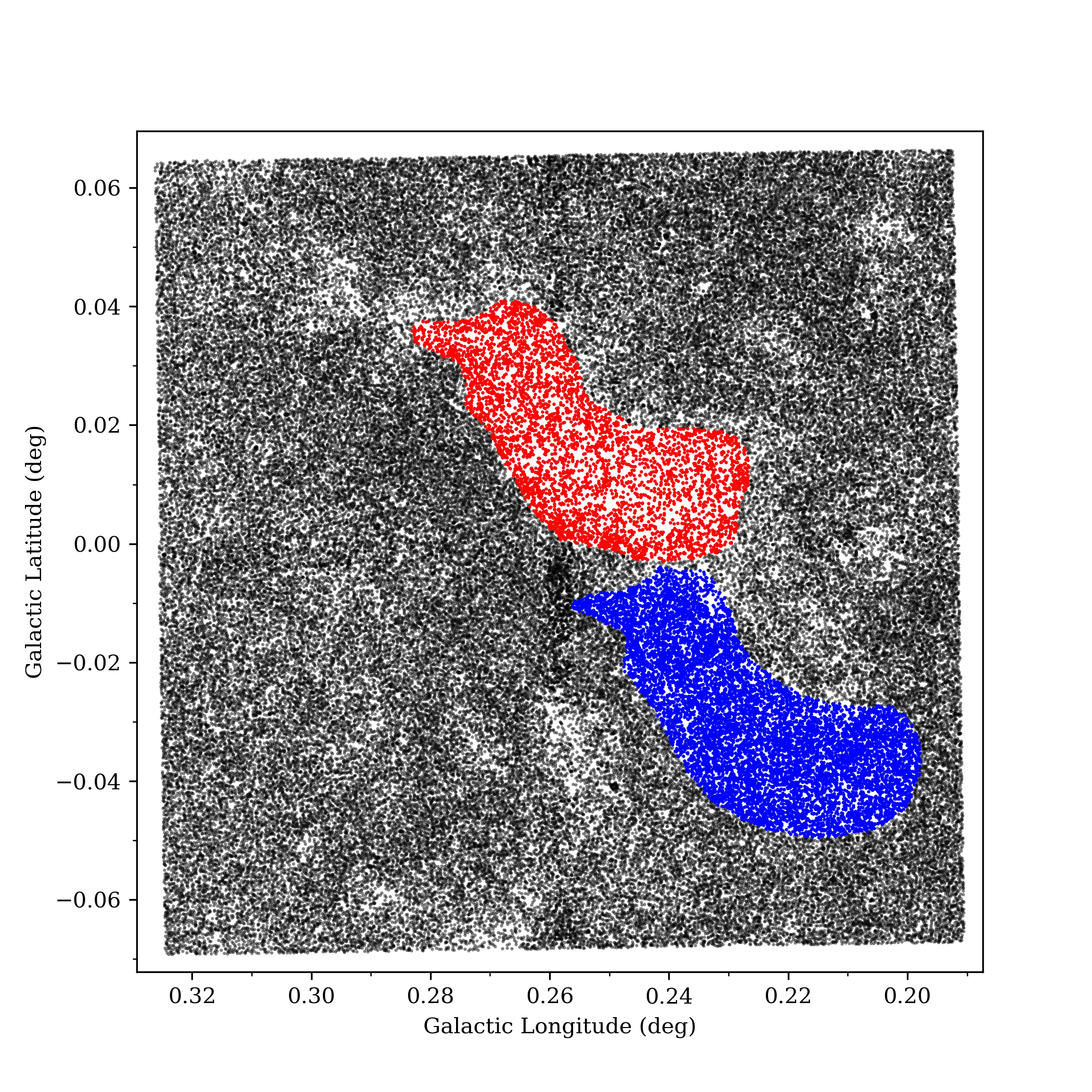}
    \caption{Spatial distribution in Galactic coordinates of all bright (i.e., K$_{\rm s}\leq$15) stars (black symbols) detected within the observed field. Stars selected within the brick (see text) are highlighted in red, while stars within a comparison control field are shown in blue.}
    \label{fig:FieldsSelection}
\end{figure}

We carried out standard photometry, including PSF modelling on each single detector from the pawprint images by using DAOPHOT \citep{daophot} and ALLFRAME \citep{allframe}. For each filter, the four single\--detector catalogues have been first cross\--matched with VVV photometry \citep{contreras-ramos+17} to perform the absolute  calibration to the 2MASS photometric system and the VVV astrometric system. 
For the stars falling in the overlapping region, the magnitude is the result of the error weighted means of the two measurements in the two adjacent detectors. An overall uncertainty of $\pm$0.05 mag in the zero\--point calibration has been estimated in both filters, while the derived astrometric solution led to rms residuals of $\leq$ 0.1 arcsec in both RA and DEC.
The final photometric catalogue listing the H and K$_{\rm s}$ magnitudes for all 203,388 detected stars was obtained through the cross\--correlation between the two single band catalogues.
The distribution of all detected stars brighter than K$_{\rm s}$=15 as a function of the Galactic coordinates is shown in Fig.\,\ref{fig:FieldsSelection} (black symbols). Bright stars within a selected region centered on the brick are marked in red, while those within a comparison control field nearby the Brick are highlighted in blue.

\section{The colour magnitude diagram}
\label{sec:cmd}

\begin{figure*}
    \centering
    \includegraphics[width=17cm]{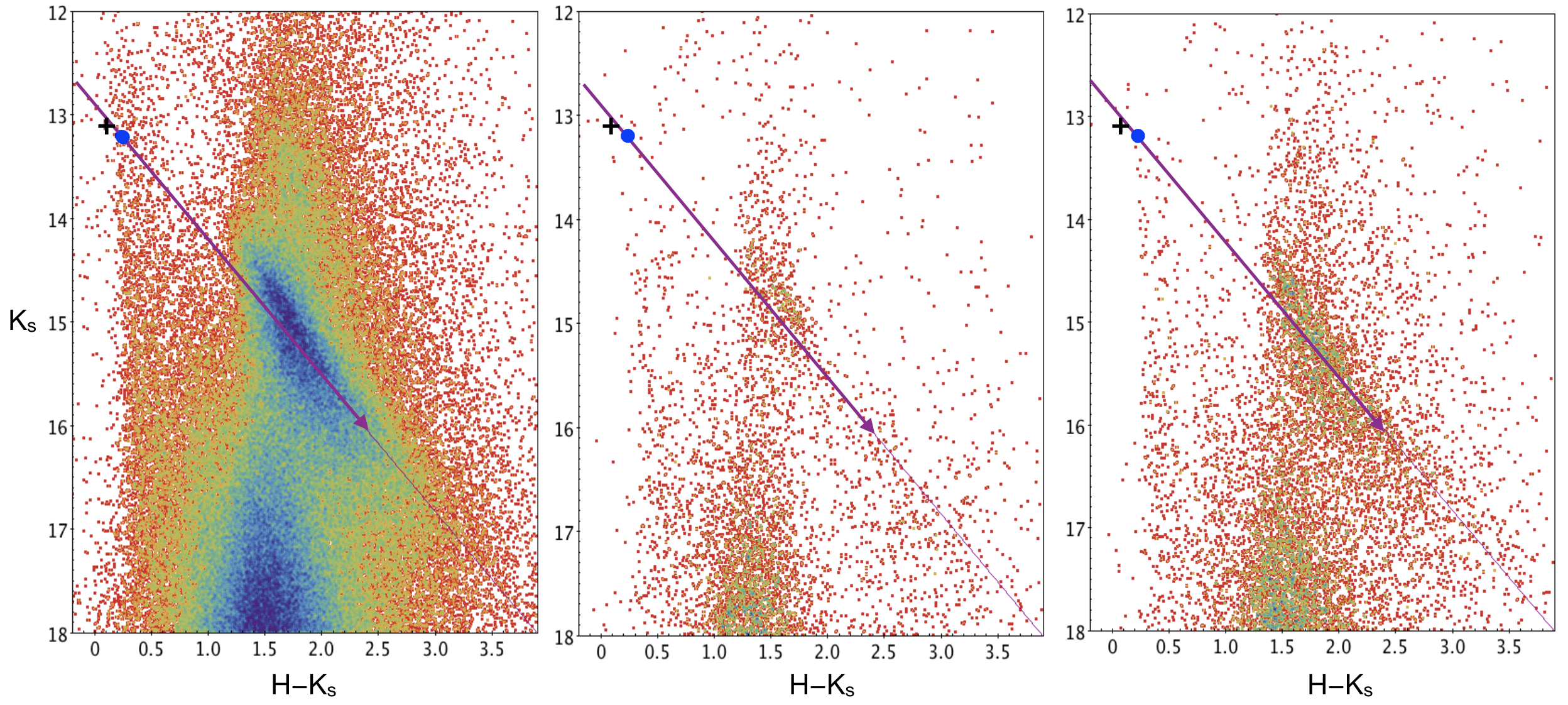}
    \caption{Left: The RC region of the CMD for all the stars detected in the HAWK-I field. Middle: CMD of the stars in front of the Brick cloud, marked in red in Fig.~\ref{fig:FieldsSelection}. Right: CMD of a field of the same size of the Brick, but outside of it, marked in blue in Fig.~\ref{fig:FieldsSelection}. In all the panels, the blue point is the location of the observed RC in Baade's Window, while the black cross is the position of the RC in an isochrone for a 12 Gyr population, with [M/H]=0.062, at a distance of 8.2 kpc. The error bar for the RC in Baade's Window, on the x-axis, is smaller than the size of the point. The purple line is the reddening vector with the slope derived by \citet{javiminniti+20} and zero point such that it passes through the RC in Baade's Window. This vector was extended to a reference high extinction typical of the Galactic center region, i.e., A$_K$=3, E(H$-$K)=2.3, corresponding to (H$-$K)=2.4.}
    \label{fig:cmd}
\end{figure*}

The colour magnitude diagram (CMD) of all the stars in the region of the red clump (RC) and red giant Branch (RGB), as detected in the HAWK-I $8'\times 8'$ mosaic field, is shown in the left panel of Fig.~\ref{fig:cmd}. 
The RC is the prominent star over-density roughly centered at K$_{\rm s}$=15, and it extends in diagonal, with the brightest stars being also bluer as they are both closer to us and affected by a lower interstellar extinction. 
The faintest RC stars, instead, are redder because they are further away and also more extincted. 
The slope of the RC in the CMD has been often used to constrain the reddening vector \citep[e.g.,]{alonso-garcia+18}, i.e., the ratio between the absolute and the relative extinction, in this case A$_{\rm Ks}$/E(H$-$K$_{\rm s}$). 
That approach, however, works only under the hypothesis that the stars are approximately all at the same distance, and therefore the only effect altering their RC magnitude is the interstellar extinction. 
In the present case, the field is located at latitude $b\sim0^\circ$ where the line\--of\--sight crosses the whole foreground disk before reaching the bulge. 
As a consequence, the distance spread is not negligible and therefore the slope of the RC in Fig.~\ref{fig:cmd} is the result of a combination of extinction and distance. In order to estimate the effect of reddening in the CMD we rely on the recent determination of the absolute\--to\--relative extinction ratio by \citet{javiminniti+20}, which is based on Classical Cepheids detected behind the bulge (at $b=0^\circ$). These authors derived a slope of A$_{\rm Ks}$= 1.308$\times$ E(H$-$K$_{\rm s}$), represented by the purple line in Fig.~\ref{fig:cmd}. 
The zero point of the purple line was constrained by using the position of the RC observed in Baade's Window, depicted as a blue point in the upper left region of the diagram. The mean colour and magnitude of the RC in Baade's Window was derived from VVV PSF photometry, and it is (H$-$K$_{\rm s}$, K$_{\rm s}$)=(0.22, 13.2). If the stellar population sampled in the present data is similar to that of Baade's Window, and it is located at the same mean distance, being only affected by a larger interstellar extinction, we should see RC stars spread along the purple line. Note that differential reddening due to the interstellar cloud density being highly non-uniform across the field, in addition to along the line of sight, also spreads stars along this reddening vector, and therefore it will not be addressed in any specific way, in what follows.

In order to double check this approach, in Fig.~\ref{fig:cmd} we also show the location of the unreddened RC as predicted by an isochrone from the BaSTI library \cite{hidalgo+18} for metallicity close to solar ([M/H]=0.06), an age of 12~Gyr and a  distance of 8.2 kpc. This point, at ((H$-$K$_{\rm s}$)$_0$, K$_{\rm s,0}$)=(0.08, 13.07) is shown as a black cross and it is extremely close to the line.
Because the RC magnitude depends very weakly on age, for ages in excess of $\sim$ 1 Gyr, the precise age adopted here does not affect the result. In other words, the cross is where the bulge RC would be expected to be, were it not reddened at all. The blue point is where the bulge RC should be, were it affected by the same reddening as Baade's Window. For more and more reddening along the line of sight, the bulge RC should be displaced to redder and fainter magnitudes, along this line, as long as its mean distance is 8.2 kpc. In fact, for colours (H$-$K$_{\rm s})>2$, and K$_{\rm s}>15$, RC stars are observed, with some spread, around the line. The spread is to be expected given the non-negligible spread in distance, around the Galactic centre. For bluer colour, however, the observed RC is significantly brighter than the purple line, especially at the blue end. Given the above mentioned low sensitivity of RC stars to age, the most plausible explanation is that the bluest RC stars are significantly closer to us, than 8.2 kpc.

The middle and right panels of Fig.~\ref{fig:cmd} show the same region of the CMD, but for stars within two sub-fields with the same shape and area, one centered on the brick (middle) and another one to the side of it, therefore in the region marked in red and blue in Fig.~\ref{fig:FieldsSelection}, respectively. 
As already evident by visual inspection of the image, in the sub-field centered on the Brick we detect fewer stars than in any sub-field of the same area not crossing the Brick. Were the Brick completely opaque, and located in the CMZ, i.e., very close to the Galactic centre, we would expect to see in front of it (middle panel) approximately half of the stars we would see to the side of it (right panel), i.e., only those in the near side of the Bulge, plus the foreground disk. Were the Brick not completely opaque, we should see even more stars than that. Instead, the middle panel contains much less than half of the stars of the right panel: between 25$\%$ and 30$\%$, depending on the selection box. This fact alone already tells us that the Brick must be much closer than the CMZ. Converting these percentages into a distance, however, is rather complicated because it requires not only a rough modeling of the distribution of stars in front and behind the Galactic centre, but also a modeling of the 3D interstellar extinction that prevents us to see all the stars right to the opposite edge of the disk.

In the absence of such a model, we decided to adopt a simpler approach, and use the magnitude difference between the observed RC in front of the Brick (middle panel) and the purple line representing stars at 8.2 kpc, affected by different amounts of reddening. This difference is $\Delta$K$_{\rm s}$=0.28 magnitudes, implying that most of the RC stars that we see in front of the Brick are located at a distance d=7.2 kpc from us, obtained by resolving:
\[ 
5 \log (8.2) - 5 \log ({\rm d}) = 0.28
\]
with respect to d.

We emphasize again that the magnitude difference of a given star from the reddening vector is a "reddening free" quantity, therefore insensitive to both absolute (along the line of sight) and differential reddening (across the field of view). 

The approach we used relies on the assumption that the HAWK-I photometry in the Brick field and the VVV photometry in Baade's Window
are directly comparable. In order to verify this, we recalculated the photometric calibration of VVV on the 2MASS system, both in Baade's Window and in the Brick field, and estimated the 1-$\sigma$ error associated to it. In Baade's Window, there are a lot of RC stars ($\sim$98,000 in one tile) in common between the VVV PSF photometry used here (in turn anchored on the CASU DR4\footnote{https://www.eso.org/rm/api/v1/public/releaseDescriptions/80}) and 2MASS. The mean offset VVV-2MASS for these stars is zero, with a negligible error as it is measured from a huge number of stars. On the contrary, in the Brick field, VVV and 2MASS have a lower number of stars in common (as the field is smaller), and they are obviously much brighter than the RC, since 2MASS only samples the upper RGB in such a crowded field. However, we did found, and correct, some offsets between the two photometric catalogues, with small differences between one VVV chip and the next (see \citet{hajdu+20} for a discussion of this problem). The statistical error on this correction was $\pm$0.03 mag.  Finally, the statistical error on the calibration between the HAWK-I instrumental photometry and the VVV one is $\pm$0.04 mag. In total, therefore, the statistical uncertainty on the distance modulus derived here is $\pm$0.05 mag, which, at a distance of 7.2 kpc, correspond to $\pm$160 pc. Possible systematics will be discussed in the next Section.

The reddening vector in Fig.~\ref{fig:cmd} was extended to (H$-$K)=2.4, as a reference for the highest extinction one
could expect close to the Galactic center, i.e., A$_K$=3. Indeed, \citet{schoedel+20} quote A$_K$=2.62 for the Nuclear Star Cluster, while \citet{javiminniti+20} quote A$_K>$3.00 for the most extincted Cepheids on the disk far side. If we redden the theoretical RC by these E(H$-$K) values, their colour would reach (H$-$K)=2.1-2.4. This means that, most likely, the RC stars at colors (H$-$K)>2.4 lie on the far side of the bulge, or even in the disk, at distances $>8.2$ kpc. It is not surprising, then, that they lie below the purple line in Fig.~\ref{fig:cmd}.

\section{Discussion and Conclusions}
\label{sec:end}

The stellar population of the Nuclear bulge is very poorly studied, due to a deadly combination of crowding, interstellar extinction and foreground disk contamination. On the contrary, the gas in the CMZ is the subject of a fairly vast literature, especially regarding the long standing problem of its low star formation rate. The latter is hard to explain, in the context of a Kennicutt-Schmidt law, given the large amount of dense molecular gas clouds that should be forming stars. Several models  have been proposed to explain the observed low star formation rate. \citet{longmore+13} and \citet{kruijssen+15} proposed that star formation might occur in bursts, triggered from the cloud passage to the pericenter of their orbit, i.e., when they come closer to the Galactic centre. More recently, \citet{armillotta+19} proposed a similar scenario, where the star formation rate has a characteristic obscillation cycle with a period of about 50 Myr, driven by feedback instabilities. The current low activity, therefore, would correspond to a temporary minimum of this cycle. 
By studying circumnuclear star formation rings in five nearby spiral galaxies, \citet{boker+08}, suggested that star formation would occur closer to the apocenter of the cloud orbits, where the ring crosses the dust lanes. 
This latter scenario is supported by the simulations presented in \citet{sormani+20}. All these works assume that the dense clouds whose star formation rate is observed to be an order of magnitude lower than expected, sit at the edge of the CMZ, at a distance between 60 and 100 pc from the Galactic centre. 

In the present paper we discuss observational evidence that one of these clouds, G0.253+0.016, a.k.a. the “Brick”, is located at 7.2 kpc from the Sun, i.e., 1 kpc closer than the Galactic centre. This result is based on the fact that RC stars seen in front of this cloud are about $\sim 0.4$ mag brighter than expected if they would lie at 8.2 kpc. They are also brighter than most of their siblings in a contiguous line of sight not crossing the cloud, and we conclude that in front of the cloud we are seeing only the stars in the foreground, thus they are brighter because they are closer to us.

An alternative interpretation could be that RC stars in front of the Brick are brighter because they are younger. The magnitude of RC stars, however, is almost completely insensitive to age, at least for ages in excess of 1 Gyr. This is the reason why they are widely used as distance indicators \citep{girardi16}. According to the latest BaSTI models \citep{hidalgo+18}, a population of stars with solar metallicity would have a RC brighter by 0.40 mag, compared to the RC of old stars, if it is as young as 480 Myr. Because the RC stars seen in front of the Brick are {\it all} this much brighter, if we want to keep the Brick in the CMZ, then we should conclude that, between the Sun and the Galactic centre there are no stars older than 480 Myr. This is obviously much less plausible than a shorter distance to the Brick.

Another alternative might be a metallicity effect. Stars in the Galactic bulge have a wide metallicity distribution approximately centered on a solar metallicity \cite[e.g.,][and references therein]{rojas-arriagada+20}. For this reason a solar metallicity isochrone was used to fix the zero point of the reddening line. The BaSTI models predict that more metal poor stars have fainter RC in K$_{\rm s}$, because they are bluer. Therefore, if the stars in front of the Brick are more metal poor than the average (hypothesis that would not be justified by any other evidence), then their RC would be fainter and therefore the Brick would be even closer to us. On the contrary, if the stars in front of the Brick have a mean metallicity twice the solar one (i.e., [Fe/H]=+0.3), then their RC magnitude would be brighter by 0.03 mag in K$_{\rm s}$. This goes in the right direction, of making the Brick closer to the Galactic centre, but only by $\sim$100 pc. This is not enough to solve the discrepancy, but it will be counted in the systematic error budget.

\begin{figure}
    \centering
    \includegraphics[width=8.5cm]{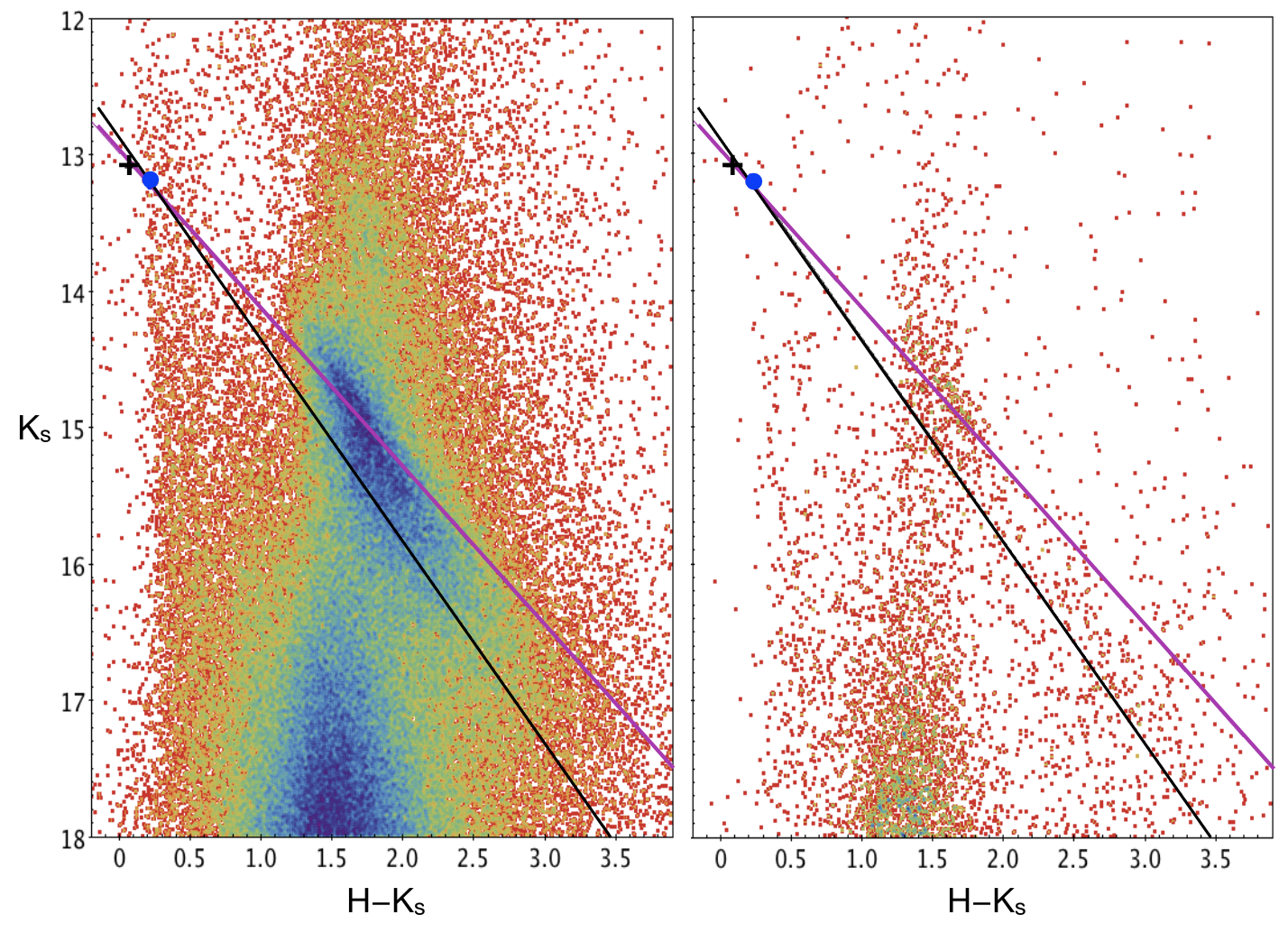}
    \caption{Same CMDs as in the left and middle panels of Fig.~\ref{fig:cmd}, with two alternative reddening vectors corresponding to the \citet{nishiyama+09} extinction law (black) and to an {\it ad hoc} vector allowing the Brick to be at the Galactic center (purple). }
    \label{fig:altredd}
\end{figure}

Finally, it might be that the reddening line drawn in Fig.~\ref{fig:cmd} is not the correct one, for this line of sight. Let us recall that we adopted this one because it was determined using classical Cepheids observed at latitude b=0$^\circ$. Cepheid variables strictly follow a period luminosity relation well studied and calibrated in several photometric bands. From the observed pulsational period, one can derive their absolute magnitudes, and therefore colours, and by comparison with the observed colours, derive the reddenings and extinction in different bands \citep{javiminniti+20}. Several other studies found that, when moving from |b|=$4^\circ$ to lower Galactic latitudes, the most appropriate extinction law is the one derived by \citet{nishiyama+09}. None of these studies, however, including \citet{nishiyama+09}, reaches a projected distance from the Galactic center smaller than 1$^\circ$. In the filters used here, the Nishiyama extinction law would have a slope of $\rm A_{Ks}= 1.48\times E(H-K_s)$, corresponding to the black line in Fig.\ref{fig:altredd}, when normalized to Baade's Window. Obviously, this is a much worse fit to the observed RC, and it would imply that the whole bulge, not only the Brick, is much closer than 8.2 kpc. We can make the exercise to draw a reddening line that would allow the Brick to be in the CMZ, i.e., one that passes just at the lower envelope of the RC stars observed in front of the Brick. That is the purple line in Fig.~\ref{fig:altredd}, with a slope of $\rm A_{Ks}= 1.158 \times E(H-K_s)$. Unfortunately, such a reddening vector would badly miss most of the other RC stars in this field, and it would lead us to conclude that the great majority of the stars in this direction is behind the Galactic centre.

Because the slope of the reddening line is obviously a source of systematics in our measurement, we propagate the error on the slope (0.05) quoted by \citet{javiminniti+20} in an error on the distance modulus, calculated by fixing the slope at the two extremes of this error bar, and re-anchoring on Baade's Window. The resulting mean distance modulus of the RC stars in front of the Brick would change by $\pm$0.05 mag.  By adding the squares of the possible metallicity systematics (0.03 mag) and the error on the reddening law, we estimate a total systematic error of $\pm$0.06 mag, corresponding to $\pm$200 pc at a distance of 7.2 kpc.

It would seem that the most plausible interpretation of the observed CMDs is that the Brick cloud lies 1 kpc closer to us than the Galactic centre.  Is it plausible to have such a dense gas cloud in this position? It would seem that it might be not so unexpected, as models of gas flows in a barred potential predict waves of densities all around the bar, that might produce peaks at $\sim 1$ kpc from the centre. Furthermore, according to \citet{henshaw+19} the Brick is "not a Brick", as its extension along the line of sight is similar to that in the plane of the sky, and therefore its density might be lower than previously thought.
Additional independent observational evidence supporting the foreground origin of the Brick is the lack of observed bright X\--ray emission, observed in other molecular clouds in the CMZ, and interpreted as the reflection of past flares from Sgr A$^{*}$ \citep{Ponti+13_XrayCMZ}.

If our interpretation is correct, then not only the position but also the orbit derived for this cloud have to be significantly revised, and with it the proposed scenarios for its absent star formation. 
This result highlights the need for the stellar and interstellar community to talk to each other, and make sure that the proposed scenarios for the structure, origin and evolution of the Nuclear bulge/CMZ are compatible with all the available observations.

A forthcoming, dedicated paper will be devoted to the study of the distribution of RC stars, both in magnitude and in color, in five fields observed with HAWK-I under the same observing program: one centered on the Galactic center, and two of them on each side, all 
at the same latitude. There is important information regarding the distance and reddening distribution of the stars in the Galactic plane and inner bulge that is engraved in the distribution of RC stars, and it is currently unexplored at these latitudes.

\section*{Data Availability}

Raw and pipeline processed images are available through the ESO archive (archive.eso.org). The catalogues will be shared on reasonable request to the corresponding author.

\section*{Acknowledgements}

Support for MZ is  provided by Fondecyt Regular 1191505, the  BASAL  CATA  Center  for Astrophysics  and Associated Technologies through grant PFB-06, and ANID - Millennium Science Initiative Project ICN12\_009, awarded to the Millenium Institute of Astrophysics (MAS). 
EV acknowledges the Excellence Cluster ORIGINS Funded by the Deutsche Forschungsgemeinschaft (DFG, German Research Foundation) under Germany's Excellence Strategy \-- EXC \-- 2094 \--390783311.
FS acknowledges financial support through the grants (AEI/FEDER, UE) AYA2017-89076-P, as well as by the Ministerio de Ciencia, Innovaci\'{o}n y Universidades (MCIU), through the State Budget and by the Consejer\'{i}a de Enconom\'{i}a, Industria, Comercio y Conocimiento of the Canary Islands Autonomous Community, through Regional Budget.
A.V.N acknowledges support from the National Agency for Research and Develpment (ANID), Scholarship Program Doctorado Nacional 2020 - 21201226.



\bibliographystyle{mnras}
\bibliography{biblio} 

\begin{thebibliography}{}
\makeatletter
\relax
\def\mn@urlcharsother{\let\do\@makeother \do\$\do\&\do\#\do\^\do\_\do\%\do\~}
\def\mn@doi{\begingroup\mn@urlcharsother \@ifnextchar [ {\mn@doi@}
  {\mn@doi@[]}}
\def\mn@doi@[#1]#2{\def\@tempa{#1}\ifx\@tempa\@empty \href
  {http://dx.doi.org/#2} {doi:#2}\else \href {http://dx.doi.org/#2} {#1}\fi
  \endgroup}
\def\mn@eprint#1#2{\mn@eprint@#1:#2::\@nil}
\def\mn@eprint@arXiv#1{\href {http://arxiv.org/abs/#1} {{\tt arXiv:#1}}}
\def\mn@eprint@dblp#1{\href {http://dblp.uni-trier.de/rec/bibtex/#1.xml}
  {dblp:#1}}
\def\mn@eprint@#1:#2:#3:#4\@nil{\def\@tempa {#1}\def\@tempb {#2}\def\@tempc
  {#3}\ifx \@tempc \@empty \let \@tempc \@tempb \let \@tempb \@tempa \fi \ifx
  \@tempb \@empty \def\@tempb {arXiv}\fi \@ifundefined
  {mn@eprint@\@tempb}{\@tempb:\@tempc}{\expandafter \expandafter \csname
  mn@eprint@\@tempb\endcsname \expandafter{\@tempc}}}

\bibitem[\protect\citeauthoryear{{Alonso-Garc{\'{\i}}a}
  et~al.,}{{Alonso-Garc{\'{\i}}a} et~al.}{2017}]{alonso-garcia+18}
{Alonso-Garc{\'{\i}}a} J.,  et~al., 2017, \mn@doi [\apjl]
  {10.3847/2041-8213/aa92c3}, \href
  {http://adsabs.harvard.edu/abs/2017ApJ...849L..13A} {849, L13}

\bibitem[\protect\citeauthoryear{{Armillotta}, {Krumholz}, {Di Teodoro}  \&
  {McClure-Griffiths}}{{Armillotta} et~al.}{2019}]{armillotta+19}
{Armillotta} L.,  {Krumholz} M.~R.,  {Di Teodoro} E.~M.,   {McClure-Griffiths}
  N.~M.,  2019, \mn@doi [\mnras] {10.1093/mnras/stz2880}, \href
  {https://ui.adsabs.harvard.edu/abs/2019MNRAS.490.4401A} {490, 4401}

\bibitem[\protect\citeauthoryear{{Barnes}, {Longmore}, {Battersby}, {Bally},
  {Kruijssen}, {Henshaw}  \& {Walker}}{{Barnes} et~al.}{2017}]{barnes+17}
{Barnes} A.~T.,  {Longmore} S.~N.,  {Battersby} C.,  {Bally} J.,  {Kruijssen}
  J.~M.~D.,  {Henshaw} J.~D.,   {Walker} D.~L.,  2017, \mn@doi [\mnras]
  {10.1093/mnras/stx941}, \href
  {https://ui.adsabs.harvard.edu/abs/2017MNRAS.469.2263B} {469, 2263}

\bibitem[\protect\citeauthoryear{{Battersby} et~al.,}{{Battersby}
  et~al.}{2020}]{battersby+20}
{Battersby} C.,  et~al., 2020, arXiv e-prints, \href
  {https://ui.adsabs.harvard.edu/abs/2020arXiv200705023B} {p. arXiv:2007.05023}

\bibitem[\protect\citeauthoryear{{Bigiel}, {Leroy}, {Blitz}, {Bolatto}, {da
  Cunha}, {Rosolowsky}, {Sandstrom}  \& {Usero}}{{Bigiel}
  et~al.}{2015}]{bigiel+15}
{Bigiel} F.,  {Leroy} A.~K.,  {Blitz} L.,  {Bolatto} A.~D.,  {da Cunha} E.,
  {Rosolowsky} E.,  {Sandstrom} K.,   {Usero} A.,  2015, \mn@doi [\apj]
  {10.1088/0004-637X/815/2/103}, \href
  {https://ui.adsabs.harvard.edu/abs/2015ApJ...815..103B} {815, 103}

\bibitem[\protect\citeauthoryear{{Bigiel} et~al.,}{{Bigiel}
  et~al.}{2016}]{bigiel+16}
{Bigiel} F.,  et~al., 2016, \mn@doi [\apjl] {10.3847/2041-8205/822/2/L26},
  \href {https://ui.adsabs.harvard.edu/abs/2016ApJ...822L..26B} {822, L26}

\bibitem[\protect\citeauthoryear{{B{\"o}ker}, {Falc{\'o}n-Barroso},
  {Schinnerer}, {Knapen}  \& {Ryder}}{{B{\"o}ker} et~al.}{2008}]{boker+08}
{B{\"o}ker} T.,  {Falc{\'o}n-Barroso} J.,  {Schinnerer} E.,  {Knapen} J.~H.,
  {Ryder} S.,  2008, \mn@doi [\aj] {10.1088/0004-6256/135/2/479}, \href
  {https://ui.adsabs.harvard.edu/abs/2008AJ....135..479B} {135, 479}

\bibitem[\protect\citeauthoryear{{Clarkson}, {Ghez}, {Morris}, {Lu}, {Stolte},
  {McCrady}, {Do}  \& {Yelda}}{{Clarkson} et~al.}{2012}]{clarkson+12}
{Clarkson} W.~I.,  {Ghez} A.~M.,  {Morris} M.~R.,  {Lu} J.~R.,  {Stolte} A.,
  {McCrady} N.,  {Do} T.,   {Yelda} S.,  2012, \mn@doi [\apj]
  {10.1088/0004-637X/751/2/132}, \href
  {https://ui.adsabs.harvard.edu/abs/2012ApJ...751..132C} {751, 132}

\bibitem[\protect\citeauthoryear{{Contreras Ramos} et~al.,}{{Contreras Ramos}
  et~al.}{2017}]{contreras-ramos+17}
{Contreras Ramos} R.,  et~al., 2017, \mn@doi [\aap]
  {10.1051/0004-6361/201731462}, \href
  {https://ui.adsabs.harvard.edu/abs/2017A&A...608A.140C} {608, A140}

\bibitem[\protect\citeauthoryear{{Dahmen}, {Huttemeister}, {Wilson}  \&
  {Mauersberger}}{{Dahmen} et~al.}{1998}]{dahmen+98}
{Dahmen} G.,  {Huttemeister} S.,  {Wilson} T.~L.,   {Mauersberger} R.,  1998,
  \aap, \href {https://ui.adsabs.harvard.edu/abs/1998A&A...331..959D} {331,
  959}

\bibitem[\protect\citeauthoryear{{Dong}, {Wang}  \& {Morris}}{{Dong}
  et~al.}{2012}]{dong+12}
{Dong} H.,  {Wang} Q.~D.,   {Morris} M.~R.,  2012, \mn@doi [\mnras]
  {10.1111/j.1365-2966.2012.21200.x}, \href
  {https://ui.adsabs.harvard.edu/abs/2012MNRAS.425..884D} {425, 884}

\bibitem[\protect\citeauthoryear{{Evans}, {Heiderman}  \&
  {Vutisalchavakul}}{{Evans} et~al.}{2014}]{evans+14}
{Evans} Neal~J. I.,  {Heiderman} A.,   {Vutisalchavakul} N.,  2014, \mn@doi
  [\apj] {10.1088/0004-637X/782/2/114}, \href
  {https://ui.adsabs.harvard.edu/abs/2014ApJ...782..114E} {782, 114}

\bibitem[\protect\citeauthoryear{{Fritz} et~al.,}{{Fritz}
  et~al.}{2016}]{fritz+16}
{Fritz} T.~K.,  et~al., 2016, \mn@doi [\apj] {10.3847/0004-637X/821/1/44},
  \href {https://ui.adsabs.harvard.edu/abs/2016ApJ...821...44F} {821, 44}

\bibitem[\protect\citeauthoryear{{Gallagher} et~al.,}{{Gallagher}
  et~al.}{2018}]{gallagher+18}
{Gallagher} M.~J.,  et~al., 2018, \mn@doi [\apj] {10.3847/1538-4357/aabad8},
  \href {https://ui.adsabs.harvard.edu/abs/2018ApJ...858...90G} {858, 90}

\bibitem[\protect\citeauthoryear{{Gao} \& {Solomon}}{{Gao} \&
  {Solomon}}{2004a}]{gao+04a}
{Gao} Y.,  {Solomon} P.~M.,  2004a, \mn@doi [\apjs] {10.1086/383003}, \href
  {https://ui.adsabs.harvard.edu/abs/2004ApJS..152...63G} {152, 63}

\bibitem[\protect\citeauthoryear{{Gao} \& {Solomon}}{{Gao} \&
  {Solomon}}{2004b}]{gao+04b}
{Gao} Y.,  {Solomon} P.~M.,  2004b, \mn@doi [\apj] {10.1086/382999}, \href
  {https://ui.adsabs.harvard.edu/abs/2004ApJ...606..271G} {606, 271}

\bibitem[\protect\citeauthoryear{{Girardi}}{{Girardi}}{2016}]{girardi16}
{Girardi} L.,  2016, \mn@doi [\araa] {10.1146/annurev-astro-081915-023354},
  \href {https://ui.adsabs.harvard.edu/abs/2016ARA&A..54...95G} {54, 95}

\bibitem[\protect\citeauthoryear{{Gravity Collaboration} et~al.,}{{Gravity
  Collaboration} et~al.}{2019}]{GravityCo+19}
{Gravity Collaboration} et~al., 2019, \mn@doi [\aap]
  {10.1051/0004-6361/201935656}, \href
  {https://ui.adsabs.harvard.edu/abs/2019A&A...625L..10G} {625, L10}

\bibitem[\protect\citeauthoryear{{Habibi}, {Stolte}  \& {Harfst}}{{Habibi}
  et~al.}{2014}]{habibi+14}
{Habibi} M.,  {Stolte} A.,   {Harfst} S.,  2014, \mn@doi [\aap]
  {10.1051/0004-6361/201323030}, \href
  {https://ui.adsabs.harvard.edu/abs/2014A&A...566A...6H} {566, A6}

\bibitem[\protect\citeauthoryear{{Hajdu}, {D{\'e}k{\'a}ny}, {Catelan}  \&
  {Grebel}}{{Hajdu} et~al.}{2020}]{hajdu+20}
{Hajdu} G.,  {D{\'e}k{\'a}ny} I.,  {Catelan} M.,   {Grebel} E.~K.,  2020,
  \mn@doi [Experimental Astronomy] {10.1007/s10686-020-09661-0}, \href
  {https://ui.adsabs.harvard.edu/abs/2020ExA....49..217H} {49, 217}

\bibitem[\protect\citeauthoryear{{Heiderman}, {Evans}, {Allen}, {Huard}  \&
  {Heyer}}{{Heiderman} et~al.}{2010}]{heiderman+10}
{Heiderman} A.,  {Evans} Neal~J. I.,  {Allen} L.~E.,  {Huard} T.,   {Heyer} M.,
   2010, \mn@doi [\apj] {10.1088/0004-637X/723/2/1019}, \href
  {https://ui.adsabs.harvard.edu/abs/2010ApJ...723.1019H} {723, 1019}

\bibitem[\protect\citeauthoryear{{Henshaw} et~al.,}{{Henshaw}
  et~al.}{2019}]{henshaw+19}
{Henshaw} J.~D.,  et~al., 2019, \mn@doi [\mnras] {10.1093/mnras/stz471}, \href
  {https://ui.adsabs.harvard.edu/abs/2019MNRAS.485.2457H} {485, 2457}

\bibitem[\protect\citeauthoryear{{Heyer} \& {Dame}}{{Heyer} \&
  {Dame}}{2015}]{heyer+15}
{Heyer} M.,  {Dame} T.~M.,  2015, \mn@doi [\araa]
  {10.1146/annurev-astro-082214-122324}, \href
  {https://ui.adsabs.harvard.edu/abs/2015ARA&A..53..583H} {53, 583}

\bibitem[\protect\citeauthoryear{Hidalgo et~al.,}{Hidalgo
  et~al.}{2018}]{hidalgo+18}
Hidalgo S.~L.,  et~al., 2018, \mn@doi [The Astrophysical Journal]
  {10.3847/1538-4357/aab158}, 856, 125

\bibitem[\protect\citeauthoryear{{Hosek}, {Lu}, {Anderson}, {Ghez}, {Morris}
  \& {Clarkson}}{{Hosek} et~al.}{2015}]{hosek+15}
{Hosek} Matthew~W. J.,  {Lu} J.~R.,  {Anderson} J.,  {Ghez} A.~M.,  {Morris}
  M.~R.,   {Clarkson} W.~I.,  2015, \mn@doi [\apj]
  {10.1088/0004-637X/813/1/27}, \href
  {https://ui.adsabs.harvard.edu/abs/2015ApJ...813...27H} {813, 27}

\bibitem[\protect\citeauthoryear{{Hu{\ss}mann}, {Stolte}, {Brandner}, {Gennaro}
   \& {Liermann}}{{Hu{\ss}mann} et~al.}{2012}]{hussmann+12}
{Hu{\ss}mann} B.,  {Stolte} A.,  {Brandner} W.,  {Gennaro} M.,   {Liermann} A.,
   2012, \mn@doi [\aap] {10.1051/0004-6361/201117637}, \href
  {https://ui.adsabs.harvard.edu/abs/2012A&A...540A..57H} {540, A57}

\bibitem[\protect\citeauthoryear{{Immer}, {Schuller}, {Omont}  \&
  {Menten}}{{Immer} et~al.}{2012}]{immer+12}
{Immer} K.,  {Schuller} F.,  {Omont} A.,   {Menten} K.~M.,  2012, \mn@doi
  [\aap] {10.1051/0004-6361/201117857}, \href
  {https://ui.adsabs.harvard.edu/abs/2012A&A...537A.121I} {537, A121}

\bibitem[\protect\citeauthoryear{{Kauffmann}, {Pillai}, {Zhang}, {Menten},
  {Goldsmith}, {Lu}  \& {Guzm{\'a}n}}{{Kauffmann} et~al.}{2017}]{kauffmann+17}
{Kauffmann} J.,  {Pillai} T.,  {Zhang} Q.,  {Menten} K.~M.,  {Goldsmith} P.~F.,
   {Lu} X.,   {Guzm{\'a}n} A.~E.,  2017, \mn@doi [\aap]
  {10.1051/0004-6361/201628088}, \href
  {https://ui.adsabs.harvard.edu/abs/2017A&A...603A..89K} {603, A89}

\bibitem[\protect\citeauthoryear{{Kennicutt}}{{Kennicutt}}{1989}]{kennicutt+89}
{Kennicutt} Robert~C. J.,  1989, \mn@doi [\apj] {10.1086/167834}, \href
  {https://ui.adsabs.harvard.edu/abs/1989ApJ...344..685K} {344, 685}

\bibitem[\protect\citeauthoryear{{Kennicutt}}{{Kennicutt}}{1998}]{kennicutt+98}
{Kennicutt} Robert~C. J.,  1998, \mn@doi [\apj] {10.1086/305588}, \href
  {https://ui.adsabs.harvard.edu/abs/1998ApJ...498..541K} {498, 541}

\bibitem[\protect\citeauthoryear{{Krips}, {Neri}, {Garc{\'\i}a-Burillo},
  {Mart{\'\i}n}, {Combes}, {Graci{\'a}-Carpio}  \& {Eckart}}{{Krips}
  et~al.}{2008}]{krips+08}
{Krips} M.,  {Neri} R.,  {Garc{\'\i}a-Burillo} S.,  {Mart{\'\i}n} S.,  {Combes}
  F.,  {Graci{\'a}-Carpio} J.,   {Eckart} A.,  2008, \mn@doi [\apj]
  {10.1086/527367}, \href
  {https://ui.adsabs.harvard.edu/abs/2008ApJ...677..262K} {677, 262}

\bibitem[\protect\citeauthoryear{{Kruijssen} \& {Longmore}}{{Kruijssen} \&
  {Longmore}}{2013}]{kruijssen+13}
{Kruijssen} J.~M.~D.,  {Longmore} S.~N.,  2013, \mn@doi [\mnras]
  {10.1093/mnras/stt1634}, \href
  {https://ui.adsabs.harvard.edu/abs/2013MNRAS.435.2598K} {435, 2598}

\bibitem[\protect\citeauthoryear{{Kruijssen}, {Dale}  \&
  {Longmore}}{{Kruijssen} et~al.}{2015}]{kruijssen+15}
{Kruijssen} J.~M.~D.,  {Dale} J.~E.,   {Longmore} S.~N.,  2015, \mn@doi
  [\mnras] {10.1093/mnras/stu2526}, \href
  {https://ui.adsabs.harvard.edu/abs/2015MNRAS.447.1059K} {447, 1059}

\bibitem[\protect\citeauthoryear{{Lada}, {Lombardi}  \& {Alves}}{{Lada}
  et~al.}{2010}]{lada+10}
{Lada} C.~J.,  {Lombardi} M.,   {Alves} J.~F.,  2010, \mn@doi [\apj]
  {10.1088/0004-637X/724/1/687}, \href
  {https://ui.adsabs.harvard.edu/abs/2010ApJ...724..687L} {724, 687}

\bibitem[\protect\citeauthoryear{{Lada}, {Forbrich}, {Lombardi}  \&
  {Alves}}{{Lada} et~al.}{2012}]{lada+12}
{Lada} C.~J.,  {Forbrich} J.,  {Lombardi} M.,   {Alves} J.~F.,  2012, \mn@doi
  [\apj] {10.1088/0004-637X/745/2/190}, \href
  {https://ui.adsabs.harvard.edu/abs/2012ApJ...745..190L} {745, 190}

\bibitem[\protect\citeauthoryear{{Launhardt}, {Zylka}  \& {Mezger}}{{Launhardt}
  et~al.}{2002}]{launhardt+02}
{Launhardt} R.,  {Zylka} R.,   {Mezger} P.~G.,  2002, \mn@doi [\aap]
  {10.1051/0004-6361:20020017}, \href
  {https://ui.adsabs.harvard.edu/abs/2002A&A...384..112L} {384, 112}

\bibitem[\protect\citeauthoryear{{Liermann}, {Hamann}  \&
  {Oskinova}}{{Liermann} et~al.}{2012}]{liermann+12}
{Liermann} A.,  {Hamann} W.~R.,   {Oskinova} L.~M.,  2012, \mn@doi [\aap]
  {10.1051/0004-6361/201117534}, \href
  {https://ui.adsabs.harvard.edu/abs/2012A&A...540A..14L} {540, A14}

\bibitem[\protect\citeauthoryear{{Longmore} et~al.,}{{Longmore}
  et~al.}{2012}]{brick-Longmore+12}
{Longmore} S.~N.,  et~al., 2012, \mn@doi [\apj] {10.1088/0004-637X/746/2/117},
  \href {https://ui.adsabs.harvard.edu/abs/2012ApJ...746..117L} {746, 117}

\bibitem[\protect\citeauthoryear{{Longmore} et~al.,}{{Longmore}
  et~al.}{2013}]{longmore+13}
{Longmore} S.~N.,  et~al., 2013, \mn@doi [\mnras] {10.1093/mnras/sts376}, \href
  {https://ui.adsabs.harvard.edu/abs/2013MNRAS.429..987L} {429, 987}

\bibitem[\protect\citeauthoryear{{Minniti} et~al.,}{{Minniti}
  et~al.}{2020}]{javiminniti+20}
{Minniti} J.~H.,  et~al., 2020, \mn@doi [\aap] {10.1051/0004-6361/202037575},
  \href {https://ui.adsabs.harvard.edu/abs/2020A&A...640A..92M} {640, A92}

\bibitem[\protect\citeauthoryear{{Molinari} et~al.,}{{Molinari}
  et~al.}{2010}]{molinari+10}
{Molinari} S.,  et~al., 2010, \mn@doi [\aap] {10.1051/0004-6361/201014659},
  \href {https://ui.adsabs.harvard.edu/abs/2010A&A...518L.100M} {518, L100}

\bibitem[\protect\citeauthoryear{{Molinari} et~al.,}{{Molinari}
  et~al.}{2011}]{molinari+11}
{Molinari} S.,  et~al., 2011, \mn@doi [\apjl] {10.1088/2041-8205/735/2/L33},
  \href {https://ui.adsabs.harvard.edu/abs/2011ApJ...735L..33M} {735, L33}

\bibitem[\protect\citeauthoryear{{Morselli} et~al.,}{{Morselli}
  et~al.}{2020}]{morselli+2020}
{Morselli} L.,  et~al., 2020, \mn@doi [\mnras] {10.1093/mnras/staa1811}, \href
  {https://ui.adsabs.harvard.edu/abs/2020MNRAS.496.4606M} {496, 4606}

\bibitem[\protect\citeauthoryear{{Nakanishi} \& {Sofue}}{{Nakanishi} \&
  {Sofue}}{2016}]{nakanishi+16}
{Nakanishi} H.,  {Sofue} Y.,  2016, \mn@doi [\pasj] {10.1093/pasj/psv108},
  \href {https://ui.adsabs.harvard.edu/abs/2016PASJ...68....5N} {68, 5}

\bibitem[\protect\citeauthoryear{{Nishiyama}, {Tamura}, {Hatano}, {Kato},
  {Tanab{\'e}}, {Sugitani}  \& {Nagata}}{{Nishiyama}
  et~al.}{2009}]{nishiyama+09}
{Nishiyama} S.,  {Tamura} M.,  {Hatano} H.,  {Kato} D.,  {Tanab{\'e}} T.,
  {Sugitani} K.,   {Nagata} T.,  2009, \mn@doi [\apj]
  {10.1088/0004-637X/696/2/1407}, \href
  {https://ui.adsabs.harvard.edu/abs/2009ApJ...696.1407N} {696, 1407}

\bibitem[\protect\citeauthoryear{{Nogueras-Lara} et~al.,}{{Nogueras-Lara}
  et~al.}{2018}]{nogueras-lara+18}
{Nogueras-Lara} F.,  et~al., 2018, \mn@doi [\aap]
  {10.1051/0004-6361/201833518}, \href
  {https://ui.adsabs.harvard.edu/abs/2018A&A...620A..83N} {620, A83}

\bibitem[\protect\citeauthoryear{{Nogueras-Lara} et~al.,}{{Nogueras-Lara}
  et~al.}{2019}]{nogueras-lara+20}
{Nogueras-Lara} F.,  et~al., 2019, \mn@doi [Nature Astronomy]
  {10.1038/s41550-019-0967-9}, \href
  {https://ui.adsabs.harvard.edu/abs/2020NatAs...4..377N} {4, 377}

\bibitem[\protect\citeauthoryear{{Pierce-Price} et~al.,}{{Pierce-Price}
  et~al.}{2000}]{pierce-price+00}
{Pierce-Price} D.,  et~al., 2000, \mn@doi [\apjl] {10.1086/317884}, \href
  {https://ui.adsabs.harvard.edu/abs/2000ApJ...545L.121P} {545, L121}

\bibitem[\protect\citeauthoryear{{Ponti}, {Morris}, {Terrier}  \&
  {Goldwurm}}{{Ponti} et~al.}{2013}]{Ponti+13_XrayCMZ}
{Ponti} G.,  {Morris} M.~R.,  {Terrier} R.,   {Goldwurm} A.,  2013, in {Torres}
  D.~F.,  {Reimer} O.,  eds, ~ Vol. 34, Cosmic Rays in Star-Forming
  Environments. p.~331 (\mn@eprint {arXiv} {1210.3034}),
  \mn@doi{10.1007/978-3-642-35410-6_26}

\bibitem[\protect\citeauthoryear{{Querejeta} et~al.,}{{Querejeta}
  et~al.}{2019}]{querejeta+19}
{Querejeta} M.,  et~al., 2019, \mn@doi [\aap] {10.1051/0004-6361/201834915},
  \href {https://ui.adsabs.harvard.edu/abs/2019A&A...625A..19Q} {625, A19}

\bibitem[\protect\citeauthoryear{{Ridley}, {Sormani}, {Tre{\ss}}, {Magorrian}
  \& {Klessen}}{{Ridley} et~al.}{2017}]{ridley+17}
{Ridley} M. G.~L.,  {Sormani} M.~C.,  {Tre{\ss}} R.~G.,  {Magorrian} J.,
  {Klessen} R.~S.,  2017, \mn@doi [\mnras] {10.1093/mnras/stx944}, \href
  {https://ui.adsabs.harvard.edu/abs/2017MNRAS.469.2251R} {469, 2251}

\bibitem[\protect\citeauthoryear{{Rojas-Arriagada} et~al.,}{{Rojas-Arriagada}
  et~al.}{2020}]{rojas-arriagada+20}
{Rojas-Arriagada} A.,  et~al., 2020, \mn@doi [\mnras] {10.1093/mnras/staa2807},
  \href {https://ui.adsabs.harvard.edu/abs/2020MNRAS.499.1037R} {499, 1037}

\bibitem[\protect\citeauthoryear{{Roman-Duval}, {Heyer}, {Brunt}, {Clark},
  {Klessen}  \& {Shetty}}{{Roman-Duval} et~al.}{2016}]{roman-duval+16}
{Roman-Duval} J.,  {Heyer} M.,  {Brunt} C.~M.,  {Clark} P.,  {Klessen} R.,
  {Shetty} R.,  2016, \mn@doi [\apj] {10.3847/0004-637X/818/2/144}, \href
  {https://ui.adsabs.harvard.edu/abs/2016ApJ...818..144R} {818, 144}

\bibitem[\protect\citeauthoryear{{Sawada}, {Hasegawa}, {Handa}  \&
  {Cohen}}{{Sawada} et~al.}{2004}]{sawada+04}
{Sawada} T.,  {Hasegawa} T.,  {Handa} T.,   {Cohen} R.~J.,  2004, \mn@doi
  [\mnras] {10.1111/j.1365-2966.2004.07603.x}, \href
  {https://ui.adsabs.harvard.edu/abs/2004MNRAS.349.1167S} {349, 1167}

\bibitem[\protect\citeauthoryear{{Sch{\"o}del}, {Nogueras-Lara},
  {Gallego-Cano}, {Shahzamanian}, {Gallego-Calvente}  \&
  {Gardini}}{{Sch{\"o}del} et~al.}{2020}]{schoedel+20}
{Sch{\"o}del} R.,  {Nogueras-Lara} F.,  {Gallego-Cano} E.,  {Shahzamanian} B.,
  {Gallego-Calvente} A.~T.,   {Gardini} A.,  2020, \mn@doi [\aap]
  {10.1051/0004-6361/201936688}, \href
  {https://ui.adsabs.harvard.edu/abs/2020A&A...641A.102S} {641, A102}

\bibitem[\protect\citeauthoryear{{Sch{\"o}nrich}, {Aumer}  \&
  {Sale}}{{Sch{\"o}nrich} et~al.}{2015}]{schoenrich+15}
{Sch{\"o}nrich} R.,  {Aumer} M.,   {Sale} S.~E.,  2015, \mn@doi [\apjl]
  {10.1088/2041-8205/812/2/L21}, \href
  {https://ui.adsabs.harvard.edu/abs/2015ApJ...812L..21S} {812, L21}

\bibitem[\protect\citeauthoryear{{Shin} \& {Kim}}{{Shin} \&
  {Kim}}{2016}]{shin+16}
{Shin} J.,  {Kim} S.~S.,  2016, \mn@doi [\mnras] {10.1093/mnras/stw1072}, \href
  {https://ui.adsabs.harvard.edu/abs/2016MNRAS.460.1854S} {460, 1854}

\bibitem[\protect\citeauthoryear{{Sofue}}{{Sofue}}{1995}]{sofue95}
{Sofue} Y.,  1995, \pasj, \href
  {https://ui.adsabs.harvard.edu/abs/1995PASJ...47..527S} {47, 527}

\bibitem[\protect\citeauthoryear{{Sormani}, {Tress}, {Glover}, {Klessen},
  {Battersby}, {Clark}, {Hatchfield}  \& {Smith}}{{Sormani}
  et~al.}{2020}]{sormani+20}
{Sormani} M.~C.,  {Tress} R.~G.,  {Glover} S. C.~O.,  {Klessen} R.~S.,
  {Battersby} C.~D.,  {Clark} P.~C.,  {Hatchfield} H.~P.,   {Smith} R.~J.,
  2020, \mn@doi [\mnras] {10.1093/mnras/staa1999}, \href
  {https://ui.adsabs.harvard.edu/abs/2020MNRAS.497.5024S} {497, 5024}

\bibitem[\protect\citeauthoryear{{Stetson}}{{Stetson}}{1987}]{daophot}
{Stetson} P.~B.,  1987, \mn@doi [\pasp] {10.1086/131977}, \href
  {http://adsabs.harvard.edu/abs/1987PASP...99..191S} {99, 191}

\bibitem[\protect\citeauthoryear{{Stetson}}{{Stetson}}{1994}]{allframe}
{Stetson} P.~B.,  1994, \mn@doi [\pasp] {10.1086/133378}, \href
  {http://adsabs.harvard.edu/abs/1994PASP..106..250S} {106, 250}

\bibitem[\protect\citeauthoryear{{Stolte}, {Brandner}, {Grebel}, {Lenzen}  \&
  {Lagrange}}{{Stolte} et~al.}{2005}]{stolte+05}
{Stolte} A.,  {Brandner} W.,  {Grebel} E.~K.,  {Lenzen} R.,   {Lagrange} A.-M.,
   2005, \mn@doi [\apjl] {10.1086/432909}, \href
  {https://ui.adsabs.harvard.edu/abs/2005ApJ...628L.113S} {628, L113}

\bibitem[\protect\citeauthoryear{{Usero} et~al.,}{{Usero}
  et~al.}{2015}]{usero+15}
{Usero} A.,  et~al., 2015, \mn@doi [\aj] {10.1088/0004-6256/150/4/115}, \href
  {https://ui.adsabs.harvard.edu/abs/2015AJ....150..115U} {150, 115}

\bibitem[\protect\citeauthoryear{{Valenti} et~al.,}{{Valenti}
  et~al.}{2016}]{valenti+16}
{Valenti} E.,  et~al., 2016, \mn@doi [\aap] {10.1051/0004-6361/201527500},
  \href {http://adsabs.harvard.edu/abs/2016A%26A...587L...6V} {587, L6}

\bibitem[\protect\citeauthoryear{{Wu}, {Evans}, {Gao}, {Solomon}, {Shirley}  \&
  {Vanden Bout}}{{Wu} et~al.}{2005}]{wu+05}
{Wu} J.,  {Evans} Neal~J. I.,  {Gao} Y.,  {Solomon} P.~M.,  {Shirley} Y.~L.,
  {Vanden Bout} P.~A.,  2005, \mn@doi [\apjl] {10.1086/499623}, \href
  {https://ui.adsabs.harvard.edu/abs/2005ApJ...635L.173W} {635, L173}

\makeatother
\end{thebibliography}


\bsp	
\label{lastpage}
\end{document}